\documentclass[twocolumn,prb,10pt,showpacs]{revtex4-1}
\usepackage[utf8]{inputenc}
\usepackage{graphicx}
\usepackage{amsmath}
\usepackage{amssymb}
\usepackage{bm}
\usepackage{hyperref}
\usepackage{color}
\usepackage{xcolor}
\usepackage{braket}

\begin{document}

\title{Renormalization of quantum dot $g$-factor in superconducting Rashba nanowires}

\author{Olesia Dmytruk, Denis Chevallier, Daniel Loss, and Jelena Klinovaja}
\affiliation{Department of Physics, University of Basel, Klingelbergstrasse 82, CH-4056 Basel, Switzerland}

\date{\today}
\begin{abstract}
We study analytically and numerically the renormalization of the $g$-factor in semiconducting Rashba nanowires (NWs), consisting of a normal and superconducting section.
If the potential  barrier between the sections is high,  a quantum dot (QD) is formed in the normal section. For harmonic (hard-wall) confinement, the effective $g$-factor of all QD levels is suppressed exponentially (power-law) in the product of the spin-orbit interaction (SOI) wavevector 
and the QD length. If the barrier between the two sections is removed, the $g$-factor of the emerging Andreev bound states  is suppressed less strongly. In the strong SOI regime and if the chemical potential is tuned to the SOI energy in both sections, the $g$-factor saturates to a universal constant. 
Remarkably, the effective $g$-factor shows a pronounced peak at the SOI energy as function of the chemical potentials.
In addition, if the SOI is uniform, the $g$-factor renormalization as a function of the chemical potential is given by  a universal dependence which is independent of the QD size. This prediction provides a powerful tool to determine experimentally whether the SOI in the whole NW is uniform and, moreover, gives direct access to the SOI strengths of the NW via $g$-factor measurements. In addition, it allows one to find the optimum position of the chemical potential for bringing the NW into the topological phase at large  magnetic fields.
\end{abstract}

\maketitle

\section{Introduction} 

Spin-orbit interaction (SOI) plays a central role in many modern condensed matter phenomena. It lies at the heart of spintronics, topological insulators, and many quantum computing platforms in semiconductors. In particular, SOI is used to manipulate spin states of quantum dots (QDs), the core of  quantum computation schemes based on spin qubits. \cite{loss1998quantum,kloeffel2013prospects} Also,  topological qubits, both Majorana fermions and parafermions,~\cite{fu2008superconducting,sato2009topological,lutchyn2010majorana,oreg2010helical,alicea2010majorana,potter2011majorana,
klinovaja2012electric,chevallier2012mutation,sticlet2012spin,halperin2012adiabatic,san2012ac,dominguez2012dynamical,terhal2012majorana,klinovaja2012transition,prada2012transport,degottardi2013majorana,thakurathi2013floquet,nadj2013proposal,nakosai2013majorana,klinovaja2013topological,braunecker2013interplay,
vazifeh2013self,pientka2013topological,maier2014majorana,poyhonen2014majorana,nadj2014observation,ruby2015end,pawlak2016probing,dmytruk2016josephson,deacon2017josephson} 
rely on the presence of strong SOI to generate  topological superconductivity. 
One of the most studied topological systems currently is a semiconducting Rashba nanowire (NW) brought into proximity with an $s$-wave superconductor (SC) and subjected to an external magnetic field. \cite{lutchyn2010majorana,oreg2010helical} Over the last years, such systems were extensively studied experimentally~\cite{mourik2012signatures,das2012zero,deng2012anomalous,churchill2013superconductor}, and a zero-bias conductance peak associated with the Majorana bound states (MBSs) was observed. However, this peak is far from being an unambiguous signature of the MBSs because other phenomena such as disorder, Kondo resonances, Andreev bound states (ABSs) or weak antilocalization, can also give rise to similar features in the conductance.~\cite{sasaki2000kondo,pikulin2012zero,liu2012zero,kells2012near,lee2012zero,rainis2013towards}
 In addition, it is challenging to measure experimentally the strength of the SOI in such nanowires, especially in the strong coupling regime in which the SOI strength can be substantially renormalized. \cite{potter2011engineering,reeg2017destructive,reeg2018metallization,antipov2018effects,woods2018effective,mikkelsen2018hybridization,de2018electric}
As a result, the range of SOI values reported in the literature is quite large. Moreover, the observation of helical gaps, opened by the interplay of SOI with the magnetic field, can be masked by Fabry-Perot oscillations as well as by additional resonances due to electron-electron interactions.\cite{nitta1997gate,stvreda2003antisymmetric,quay2010observation,nadj2012spectroscopy,sadreev2013effect,rainis2014conductance,zyuzin2014nuclear,sasaki2014direct,van2015spin,albrecht2016exponential,scherubl2016electrical,heedt2017signatures,estrada2018split,khrapai2018spin}
Weak-antilocalization measurements require many subbands to be filled and do not give direct access to the SOI strength of the lowest subband. Thus, there is an urgent need to identify additional ways to access and measure the SOI in such NWs. In addition, the optimal regime to observe MBSs is to tune the chemical potential to the SOI energy. However, so far, there is no reliable way known to test such a `sweet-spot' tuning experimentally.

As one eventually strives to move from the observation of zero-bias peaks associated with MBSs to assembling topological quantum computing schemes, one needs to allow for  coupling between MBSs and quantum dots.  \cite{liu2011detecting,leijnse2011scheme,golub2011kondo,zocher2013modulation,gong2014detection,leijnse2014thermoelectric,vernek2014subtle,schrade2015proximity,deng2016majorana,szombati2016josephson,ricco2016decay,hoffman2017spin,escribano2017interaction,prada2017measuring,ptok2017controlling,kobialka2018electrostatical}
Through this coupling  missing quantum gate operations (not achievable by braiding alone) can be implemented. \cite{hoffman2016universal,plugge2016roadmap,karzig2017scalable} Such systems consisting of NWs with two sections, one in the normal state hosting a QD and one in the superconducting state due to proximity-induced superconductivity, have  recently  been assembled successfully.~\cite{deng2016majorana,de2018electric} Focusing on the properties of the states localized on the QD, several experimental groups have shown that the effective $g$-factor of such localized QD states can be tuned, for example, by changing the local chemical potentials on the QD or the gate voltage on the interface between the normal and superconducting sections.\cite{kammhuber2016conductance,vaitiekenas2017effective,de2018electric,candido2018blurring}
On the other hand, from previous work on spin qubits\cite{pryor2006lande,fasth2007direct,van2012g,maier2013tunable,brauns2016electric}
studying properties of the lowest level of a normal state QD with parabolic confinement, we know that the effective $g$-factor is exponentially suppressed in the SOI strength.\cite{trif2008spin} However, there are no systematic studies of such $g$-factor renormalization for higher orbital states and, in particular, for ABSs in a QD coupled to a superconductor, arising from the interplay of SOI with the size of the QD, the applied gate voltages, the dot confinement, and other system parameters. Not surprisingly, this renormalization behavior turns out to be very rich and it is the goal of this work  to fill the gap of our understanding of such phenomena.

\begin{figure}[t]
    \includegraphics[width=8.5cm]{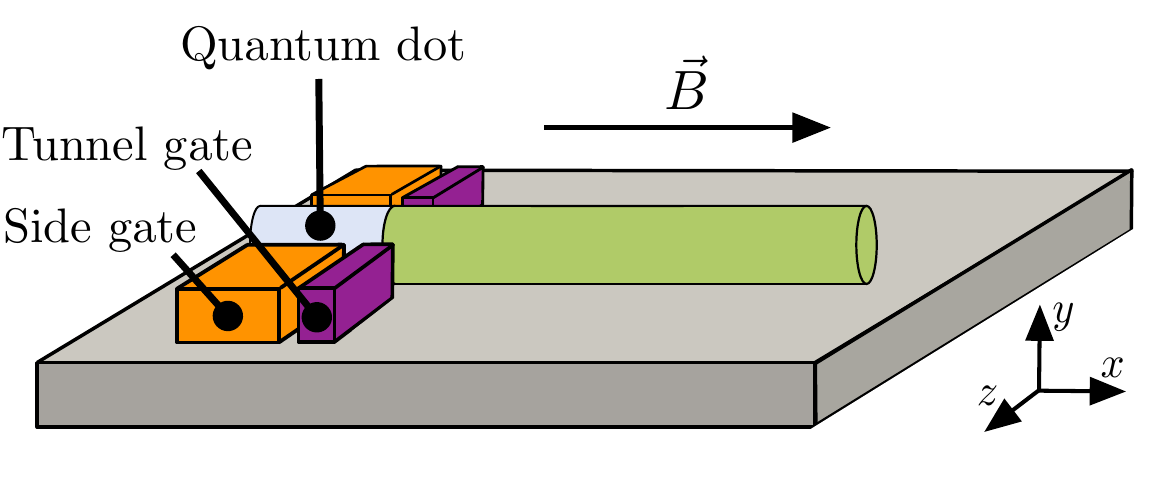}
    \caption{Schematic setup of the system: a one-dimensional  NW is aligned along the $x$-axis in the presence of an external magnetic field ${\bf B}$ applied parallel to the NW and perpendicular to the Rashba SOI vector, which points along the $z$ direction. The NW consists of two sections: In the normal section (blue), hosting a QD, the chemical potential is controlled by the external side gate (orange). The tunnel gate (violet) controls the height of the barrier potential $\mu_b$ and separates the normal section from the superconducting section (green) with proximity-induced superconductivity.}
    \label{fig:Fig1}
\end{figure}

First we investigate the behavior of the QD in the normal state. We study the $g$-factor renormalization of  Zeeman-split spin levels both numerically and analytically for two generic types of confinement. We show that for harmonic confinement the $g$-factor of ground and excited states is suppressed exponentially in the product of the SOI wavevector and the QD length, while for hard-wall confinement this suppression follows an inverse power-law. The Zeeman splitting can be even tuned to zero by changing the QD size at fixed magnetic field. By studying such zeroes, one can determine which orbital levels of the QDs are involved. As the QD is brought into contact with the superconducting section of the NW, the  suppression of the $g$-factor of the resulting ABSs gets less pronounced. In contrast to the normal-state QD, in the strong SOI regime, the $g$-factor saturates at a finite universal value, instead of being renormalized to zero in the limit of large dots. In addition, the $g$-factor is sensitive to the position of chemical potentials in both sections. Most remarkably, the $g$-factor reaches its maximum if the chemical potential is tuned to the SOI energy, exactly in the regime that is most optimal for generating the topological phase. In addition, if the SOI is uniform in the entire NW, the $g$-factor dependence on the chemical potential is given by a universal curve independent  of the QD size. This is not the case, however, if the SOI is different in two sections. In this case, to achieve the maximum value of the $g$-factor, one needs to tune the chemical potential in each section to the corresponding SOI energy. All these tunable features of the $g$-factor open up  powerful ways to access the strength of the SOI by local measurements of the $g$-factor in standard transport experiments. In addition, via $g$-factor measurements one can test whether the SOI is uniform in the NW or if it is enhanced on the QD, for example, due to the electric fields generated by local gates. 
Since $g$-factor measurements for ABSs in such NWs with QDs have already been performed in pioneering experiments by Deng et al. (see Ref. \onlinecite{deng2016majorana}), we believe that 
our predictions can be readily tested experimentally in such or similar devices.

The paper is organized as follows. In Sec.~\ref{Parabolic/sharp}, we present a study of the QD located entirely in the normal section for different types of confinement potential. We derive analytical expressions for the $g$-factor renormalization and confirm our results also numerically. In Sec.~\ref{evolution}, normal and superconducting sections are coupled such that ABSs have support in both sections. We study the dependence of the renormalized $g$-factor on system parameters such as the SOI strength, the position of the chemical potential, and the strength of proximity-induced superconductivity. In several regimes, we also provide analytical solutions, detailed derivations of which are given in the Appendix \ref{sec:zeeman}. Finally, we conclude and give some perspectives in Sec.~\ref{conclusion}.

\section{$g$-factor renormalization in isolated normal QD}\label{Parabolic/sharp}
To begin with, we focus on an isolated normal QD (without contact to the superconducting section) created by two types of confinement:  hard-wall  and  parabolic confinement potentials. The QD is hosted in a one-dimensional Rashba NW aligned along the $x$-axis in the presence of an external magnetic field applied perpendicular to the SOI vector, say,  in the $x$ direction (see  Fig.~\ref{fig:Fig1}).  The tight-binding Hamiltonian describing this system reads\cite{chevallier2016tomography} 
\begin{align}\label{eq:hamil}
&H_n = \sum_{\sigma,\sigma'}\sum_{ j=1}^{N_n-1} 
c^\dag_{j+1,\sigma}\Big[i \overline{\alpha} \sigma^y_{\sigma\sigma'}-t \delta_{\sigma\sigma'} \Big]c_{j,\sigma'}\notag\\
&\hspace{30pt}- \sum_{\sigma,\sigma'}\sum_{j=1}^{N_n} c^\dag_{j,\sigma}\Big[\mu_j \delta_{\sigma\sigma'}
- V_Z \sigma^x_{\sigma\sigma'}\Big]c_{j,\sigma'}
+{\rm H.c.},
\end{align} 
where $c_{j,\sigma}^\dag (c_{j,\sigma})$ is the creation (annihilation) operator acting on an electron with spin $\sigma$ located at site $j$ of the chain consisting of $N_n$ sites.  Here, $t = \hbar^2/\left(2 m a^2\right)$ is the hopping amplitude with $m$ being the effective mass, $a$  the lattice constant, and $\hbar$ the Planck constant (divided by $2\pi$). The spin-flip hopping amplitude $\overline{\alpha}$ is related to the Rashba SOI parameter $\alpha$ as $\overline{\alpha} = \alpha/2a$. The Zeeman energy $V_Z = g\mu_B B/2$ is determined by the $g$-factor $g$, by the Bohr magneton  $\mu_B$, and by the applied magnetic field  $B$.  The parameter $\mu_j$ denotes the chemical potential at site $j$. Via exact diagonalization of the Hamiltonian Eq.~(\ref{eq:hamil}), we can extract the spectrum of the states localized at the QD.

\begin{figure}[t!] 
\includegraphics[width=8.7cm]{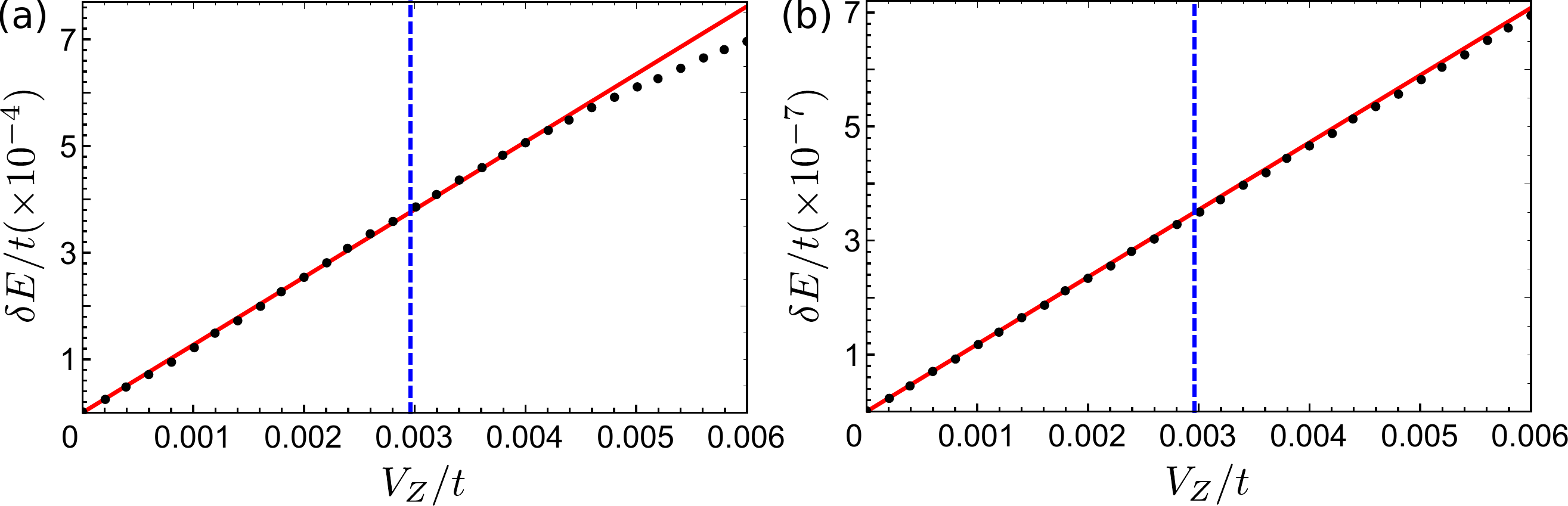}
\caption{Effective Zeeman splitting $\delta E/t$ between the two lowest energy states  for a normal QD with hard-wall confinement  as a function of Zeeman energy $V_Z/t$ (a) in the regime of weak SOI, $\overline{\alpha}/t = 0.05$, and (b) in the regime of strong SOI, $\overline{\alpha}/t = 0.4$.
The results are obtained numerically (black dotted lines) by exact diagonalization of Eq.~(\ref{eq:hamil}). The Zeeman splitting is linearly proportional to the Zeeman energy for small values of $V_Z$ that do not exceed the QD level spacing $\delta_{n=1}$ (blue dashed line).
The red solid line corresponds to the linear fit giving the renormalized values for the $g$-factor: (a) $g^*/g = 0.13$ and (b) $g^*/g = 1.2 \times 10^{-4}$.
The parameters are chosen as $N_n=100$ and $\mu_j/t = -2$. } 
\label{fig:Fig10}
\end{figure}

For the analytic treatment we consider the corresponding continuum model, with the QD being described by the effective Hamiltonian
 $H_0 = H_{kin} + H_{so} + H_Z + V$, where
the kinetic term is given by
\begin{align}
H_{kin} = -\dfrac{\hbar^2}{2m}\sum_{\sigma}\int dx\ \psi^\dag_\sigma (x)\partial^2_x \psi_{\sigma}(x),
\label{eq:kinetic}
\end{align}
and the SOI term by
\begin{align}
H_{so} = -i\alpha\sum_{\sigma,\sigma'}\int dx\ \psi^\dag_\sigma (x)\left(\sigma_z\right)_{\sigma\sigma'}\partial_x\psi_{\sigma'}(x).
\label{eq:soi}
\end{align}
The Zeeman term reads
\begin{align}
H_Z = V_Z\sum_{\sigma,\sigma'}\int dx\ \psi^\dag_\sigma (x)\left(\sigma_x\right)_{\sigma\sigma'}\psi_{\sigma'}(x),
\label{eq:zeeman}
\end{align}
and the confinement potential is given by
\begin{align}
V=\sum_{\sigma}\int dx\ \psi^\dag_\sigma (x) \mathcal{V}(x) \psi_{\sigma}(x).
\end{align}
Here, 
$\psi^\dag_\sigma (x) \left[\psi_\sigma (x)\right]$ is the creation (annihilation) operator of an electron at position $x$ with spin $\sigma/2 = \pm 1/2$, and $\sigma_{x,y,z}$ are the Pauli matrices acting on the spin of the electron.

\subsection{Hard-wall confinement potential}
First we consider a QD with  hard-wall confinement realized by tuning external gates such that the QD is isolated from the superconducting section (see Fig.~\ref{fig:Fig1}). In the absence of magnetic fields, each QD level is twofold degenerate due to Kramers degeneracy. At weak fields, the Zeeman splitting 
 $\delta E$ between these two levels is a linear function of $V_Z$, see Fig.~\ref{fig:Fig10}. We note that here we work with a single-particle Hamiltonian and the position of the chemical potential does not play a role as our QD levels are labeled from the bottom of the band. For simplicity, we fix the homogeneous chemical potential in the tight-binding model [see Eq.~(\ref{eq:hamil})] as $\mu_j=-2t$, which corresponds to hard-wall confinement.
 From  Fig.~\ref{fig:Fig10}, we conclude that for Zeeman energies being smaller than the level spacing, the Zeeman splitting $\delta E$ is linear in $V_Z$. This allows us to define the effective $g$-factor as $g^*/g = \delta E / V_Z$.

\begin{figure}[t] 
\includegraphics[width=8.7cm]{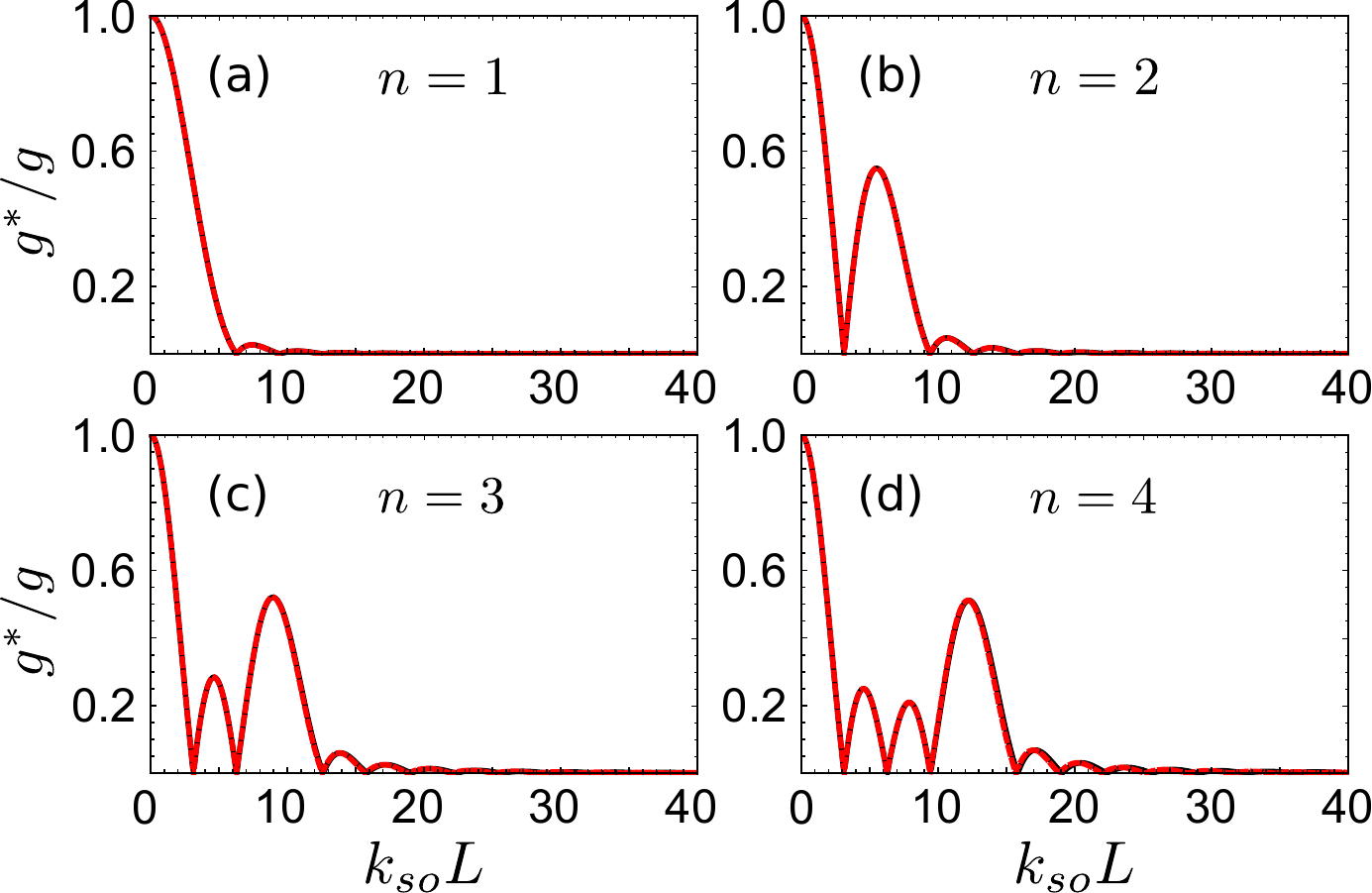}
\caption{
Renormalized $g$-factor $g^*/g$ as a function of $k_{so} L$ for a QD with  hard-wall confinement for the four lowest levels (a) $n=1$, (b)  $n=2$, (c) $n=3$, and (d) $n=4$. Here, $L$ is the length of the QD and $k_{so}=m\alpha/\hbar^2$  the SOI wavevector. The stronger the SOI or the longer the QD is, the more pronounced is the $g$-factor suppression. We note that the $g$-factor has zeroes.
The results are obtained numerically (black solid line) by exact diagonalization of Eq.~(\ref{eq:hamil}) and analytically (red dashed line) by making use of  Eq.~(\ref{eq:splittingn}). There is excellent agreement between the two approaches. 
The parameters are chosen as $N_n = 99$ and $V_Z/t=10^{-4}$.
} 
\label{fig:Fig2b}
\end{figure}

Since $\delta E$ is linear in the magnetic field we can use perturbation theory to calculate the effective $g$-factor for small values of $V_Z$. In order to extract an analytical expression of the Zeeman splitting, we first find the wavefunctions of the QD in the absence of the magnetic field and then calculate the Zeeman splitting between two energy states perturbatively. To model the hard-wall potential, for the QD of length $L = (N_n + 1)a$, we use $\mathcal{V}(x)=0$. We note that in the absence of magnetic fields, the SOI term can be effectively eliminated by performing a spin-dependent gauge transformation~\cite{braunecker2010spin,klinovaja2012composite}
\begin{align}
\psi_\sigma(x) = e^{-i\sigma k_{so}x}\tilde{\psi}_\sigma(x),
\label{eq:spingaugetransform}
\end{align}
where the new wavefunctions $\tilde{\psi}_\sigma(x)$ also satisfy vanishing boundary conditions.
Thus, the spectrum of the QD is given by $\epsilon_n = -E_{so} + n^2 \pi^2 \hbar^2/2m L^2$, where the SOI wavevector (energy) is defined as $k_{so} = m \alpha/\hbar^2$ ($E_{so}=\hbar^2 k_{so}^2/2m$), and where $n$ is a positive integer. Each energy level 
$\epsilon_n $ is twofold degenerate. The level spacing $\delta_n = \epsilon_{n+1} - \epsilon_n$ does not depend on the SOI and is given by $\delta_n = (2n+1)\pi^2 \hbar^2/2m L^2$.

The normalized wavefunctions for the $n$-th level in the absence of  magnetic fields are given by
\begin{align}
&\phi^{(n)}_{1}(x) = \sqrt{\dfrac{2}{L}}\sin\left(\dfrac{\pi n x}{L}\right)
\begin{pmatrix}
e^{-i k_{so} x}\\
0
\end{pmatrix},\\
&\phi^{(n)}_{\bar 1}(x) = \sqrt{\dfrac{2}{L}}\sin\left(\dfrac{\pi n x}{L}\right)
\begin{pmatrix}
0\\
e^{i k_{so} x}
\end{pmatrix}
.
\label{eq:wavefunction}
\end{align}
We calculate the Zeeman splitting between two spin-degenerate states $\phi^{(n)}_{1,\bar 1}$ by  using  degenerate perturbation theory in the regime $V_Z \ll \delta_n$. 
As a result, the renormalized $g$-factor of the $n$-th level is given by a compact formula,
\begin{align}
\left(g^*/g\right)_n= \left|\dfrac{\pi^2 n^2   \sin(k_{so} L)}{k_{so }L\left[\pi^2 n^2 - (k_{so} L)^2 \right]}\right|.
\label{eq:splittingn}
\end{align}

Generally, we note that the suppression of  $g^*$ is strongest for low-energy levels. In addition, we note that the degree of the renormalization is determined solely by the ratio $ L/l_{so}$, where  $L$ is the QD  length and  $l_{so}=1/k_{so}$ the  SOI length. Thus, the smaller the SOI length is, the weaker the effect of the magnetic field on the spin states gets. Generally, the suppression is governed by a power-law in $k_{so} L$, see Eq. (\ref{eq:splittingn}). In the regime of strong SOI with $l_{so}\ll L$, the magnetic field gets averaged out to zero. The physical interpretation of this effect is that the spin of the localized state effectively rotates many times over the QD size, and thereby the effect of the magnetic field applied in a given direction gets averaged out.
In addition, there are well-pronounced oscillations such that the $g$-factor periodically goes to zero.
If $k_{so}L = \pi m$, where $m$ is an integer not equal to the level number $n$, $m\neq n$, the effective $g$-factor $g^*$ vanishes. This means that if the length of the QD is chosen such that $k_{so}L = \pi m$, all levels except the $m$th level have $g^*=0$, whereas the $m$th level is finite and given by $g^*/g=1/2$. Quite remarkably, this  value  $g^*/g=1/2$ does not depend on any system parameters and is universal for all quantum dot levels. This means that, for these sets of parameters, the QD spin states of the $n$th level with $n\neq m$  are not  split by the magnetic field and stay pinned to their initial values $\epsilon_m$. 

To check the analytical expression, Eq.~(\ref{eq:splittingn}), we  determine the renormalized $g$-factor for the four lowest levels of the QD also in the tight-binding model numerically,  see Fig.~\ref{fig:Fig2b}. The numerical (black solid line) and analytical (red dashed line) results match perfectly in the regime of linear response in the $B$-field.

\begin{figure}[t] 
\includegraphics[width=8.7cm]{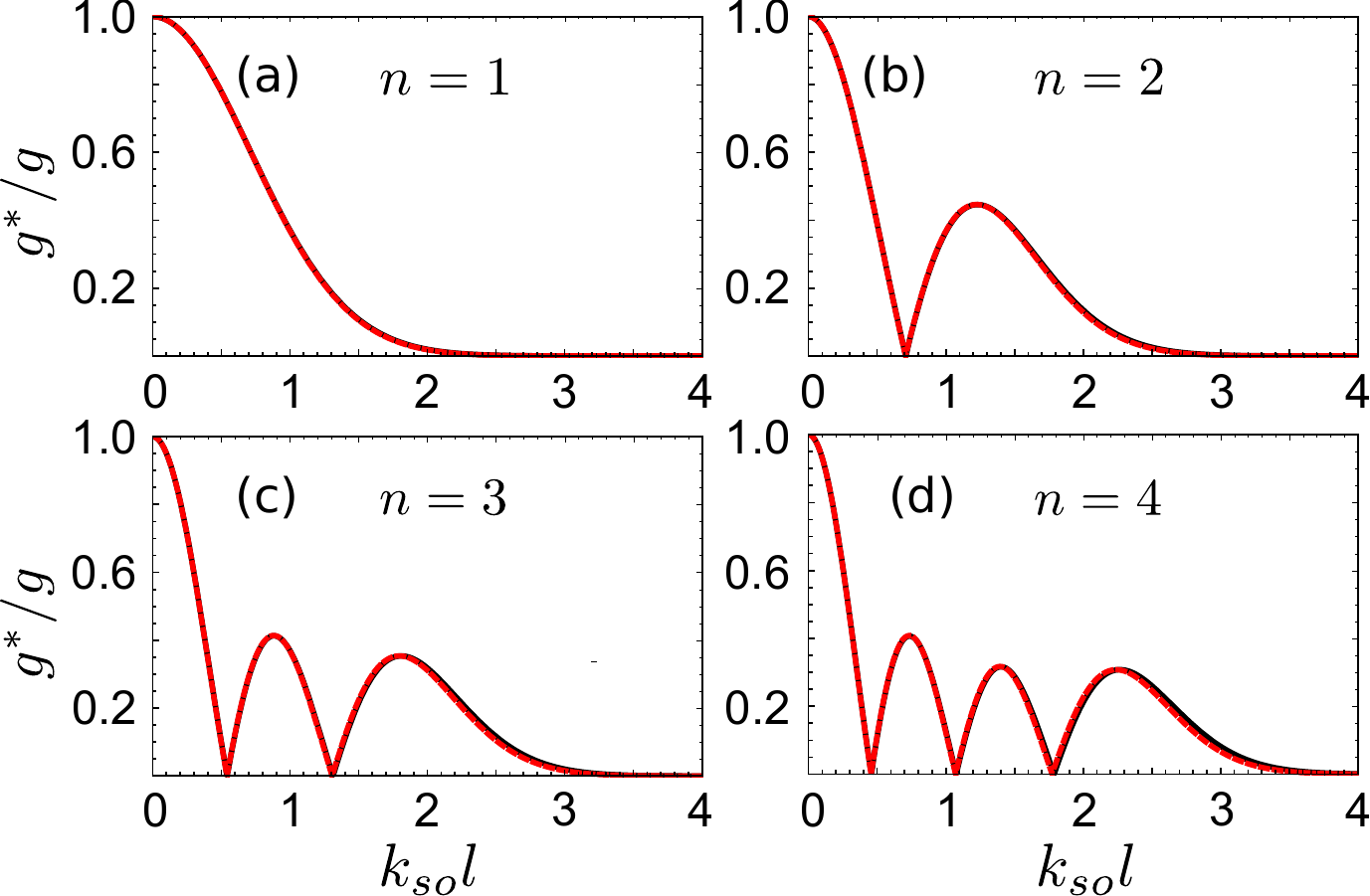}
\caption{Renormalized $g$-factor $g^*/g$ as a function of $k_{so} l$  for a QD with a harmonic confinement potential for the same set of parameters as in Fig.~\ref{fig:Fig2b}. Here, $l$ is the effective QD size and $k_{so}$ is the SOI wavevector. Again, there is an excellent agreement between numerical (black solid line) and analytical (red dashed line) results. The renormalized $g$-factor is suppressed exponentially. For higher orbital levels, the oscillations in $g^*$ allows one to fine-tune to the regime in which $g^*=0$. Numerical results are obtained for  $\hbar\omega_0 / t = 0.02$.
} 
\label{fig:Fig2a}
\end{figure}

\subsection{Harmonic confinement potential}
Next, we consider a QD created by the harmonic confinement potential given by $\mathcal{V}(x)=  m \omega_0^2 x^2 / 2$, where $l = \sqrt{\hbar / m\omega_0}$ is the effective QD size.  The quantized energy levels of such a QD are again twofold degenerate Kramers doublets  and given by \cite{liu2018spin}  
 $\epsilon_n = -E_{so} + \hbar\omega_0 (n-1/2) $,
with $n$ being a positive integer. The level spacing $\delta$ is uniform and is given by $\hbar\omega_0$.
In the absence of the magnetic field, the normalized wavefunctions for the $n$-th level read
\begin{align}
&\phi^{(n)}_1(x) = 
f_n(x)\begin{pmatrix}
e^{-i k_{so} x}\\
0
\end{pmatrix},\\
&\phi^{(n)}_{\bar 1}(x) = 
f_n(x)\begin{pmatrix}
0\\
e^{i k_{so} x}
\end{pmatrix},
\label{eq:wavefunction}
\end{align}
with
\begin{equation}\label{fnfunction}
f_n(x) = \dfrac{1}{\pi^{1/4}\sqrt{l}\sqrt{2^{n-1} \left(n-1\right)!}} e^{-x^2/2l^2} H_{n-1}(x/l).
\end{equation}
Here, $H_{n}(y)$ stands for the $n$th Hermite polynomial and is given by 
%$H_{n-1}(y) = (-1)^{n-1} e^{y^2}\dfrac{d^{n-1}}{dy^{n-1}}\left(e^{-y^2}\right)$.
$H_{n}(y) = (-1)^{n} e^{y^2}\dfrac{d^{n}}{dy^{n}}\, e^{-y^2}$.
Following the same procedure as for the hard-wall confinement we calculate the Zeeman splitting perturbatively for $V_Z \ll \hbar\omega_0$.
After straightforward calculations, we arrive at the renormalized $g$-factor of the $n$th level, 
\begin{equation}
\left(g^*/g\right)_{n}=\int_{-\infty}^{+\infty}dx\ e^{2 i k_{so}  x}f_{n}^2(x).
\end{equation}
For the lowest four orbital levels, one gets 
\begin{align}
\left(g^*/g\right)_{n} = e^{-(k_{so} l)^2}\left|\varphi_n(x)\right|,
\end{align}
where $\varphi_1(x) = 1$, $\varphi_2(x) = 1 - 2 \left(k_{so}l\right)^2$, $\varphi_3(x) = 1 -4 \left(k_{so}l\right)^2 + 2 \left(k_{so}l\right)^4$, and $\varphi_4(x) = 1 - 6 \left(k_{so}l\right)^2 + 6 \left(k_{so}l\right)^4  - 4 \left(k_{so}l\right)^6/3 $. Again we find that $\left(g^*/g\right)_{n}$ depends on the product of the SOI vector $k_{so}$ and the effective QD size $l$. In contrast to the hard-wall confinement, the $g$-factor is suppressed exponentially in $k_{so}l$, however, again, the renormalization is weaker for higher levels. With the exception of the lowest level, $g^*$ oscillates and again can be fine-tuned to zero. In total, there are $n-1$ zeroes in the $g$-factor of the $n$-th dot level.
We note that for $n=1$ we reproduce the result obtained previously \cite{trif2008spin}  by using a Schrieffer-Wolff transformation.

To support our analytical results, we compute the renormalized $g$-factor for the lowest levels also numerically, see Fig.~\ref{fig:Fig2a}. Again, there is excellent agreement between our numerical and analytical results.  Here, we have implemented a harmonic potential in the tight-binding model by working with  the chemical potential $\mu_j = -2t - \mathcal{V}_j$, where $\mathcal{V}_j = m\omega_0^2 \left(j - j_0\right)^2/2$
%$\mathcal{V}_j = \left(\hbar\omega_0\right)^2 \left(j - 1 -\left(N_n - 1\right)/2\right)^2/4t$
with $j_0 = \left(N_n + 1\right)/2$ being the center of the QD.

\section{$g$-factor renormalization of ABS levels in the QD}\label{evolution}

The barrier, separating the two NW sections, controls the strength of the superconducting pairing correlations induced on the dot levels. Above we have studied the effective $g$-factor in the regime in which the QD is separated from the superconducting section by a high potential barrier and, thus, can be considered to be localized fully in the normal section. Now, we lower this barrier such that the QD gets coupled to the superconducting section and ABS levels are induced in the QD.

The tight-binding Hamiltonian describing the barrier ($H_b$) and the superconducting section ($H_s$) are given by \cite{rainis2013towards}
\begin{align}\label{eq:hamilbarr}
&H_b = 
-\mu_b \sum_{\sigma}\sum_{j=N_n-W_b+1}^{N_n} c^\dag_{j,\sigma} 
c_{j,\sigma},\nonumber\\
&H_s = \sum_{\sigma,\sigma'}\sum_{ j=N_n+1}^{N-1} 
c^\dag_{j+1,\sigma}\Big[i \overline{\alpha} \sigma^y_{\sigma\sigma'}-t \delta_{\sigma\sigma'} \Big]c_{j,\sigma'}\notag\\
&- \sum_{\sigma,\sigma'}\sum_{j=N_n+1}^{N} c^\dag_{j,\sigma}\Big[\mu_s \delta_{\sigma\sigma'}
- V_Z \sigma^x_{\sigma\sigma'}\Big]c_{j,\sigma'}\notag\\
&-\Delta \sum_{j=N_n + 1}^{N} c_{j,\uparrow}c_{j,\downarrow}+{\rm H.c.},
\end{align} 
where $\mu_b$ is the height of the potential barrier at the interface between the normal and the superconducting sections. By changing the width of the barrier $W_b$, one controls the properties of the ABS levels localized in the QD. Here, $N$ is the total number of sites in the system and $N_s = N-N_n$ is the number of sites in the superconducting section. The superconducting pairing potential $\Delta$ is non-zero only in the superconducting section where the chemical potential is fixed to $\mu_s$.

In our considerations, we assume that the superconducting section is long enough such that the QD states are well-localized at the interface with the normal section and do not reach the right end of the NW\cite{chevallier2012mutation,cheng2012josephson,san2012ac,chevallier2013effects,crepin2014signatures,thakurathi2015majorana}.  
For simplicity, in what follows, we keep the length of the superconducting section fixed by choosing $N_s=300$ and only change the size of the QD. An external gate (shown in violet in Fig.~\ref{fig:Fig1}) allows one to tune the height of the potential barrier at the interface. 

\begin{figure}[t] 
\includegraphics[width=8cm]{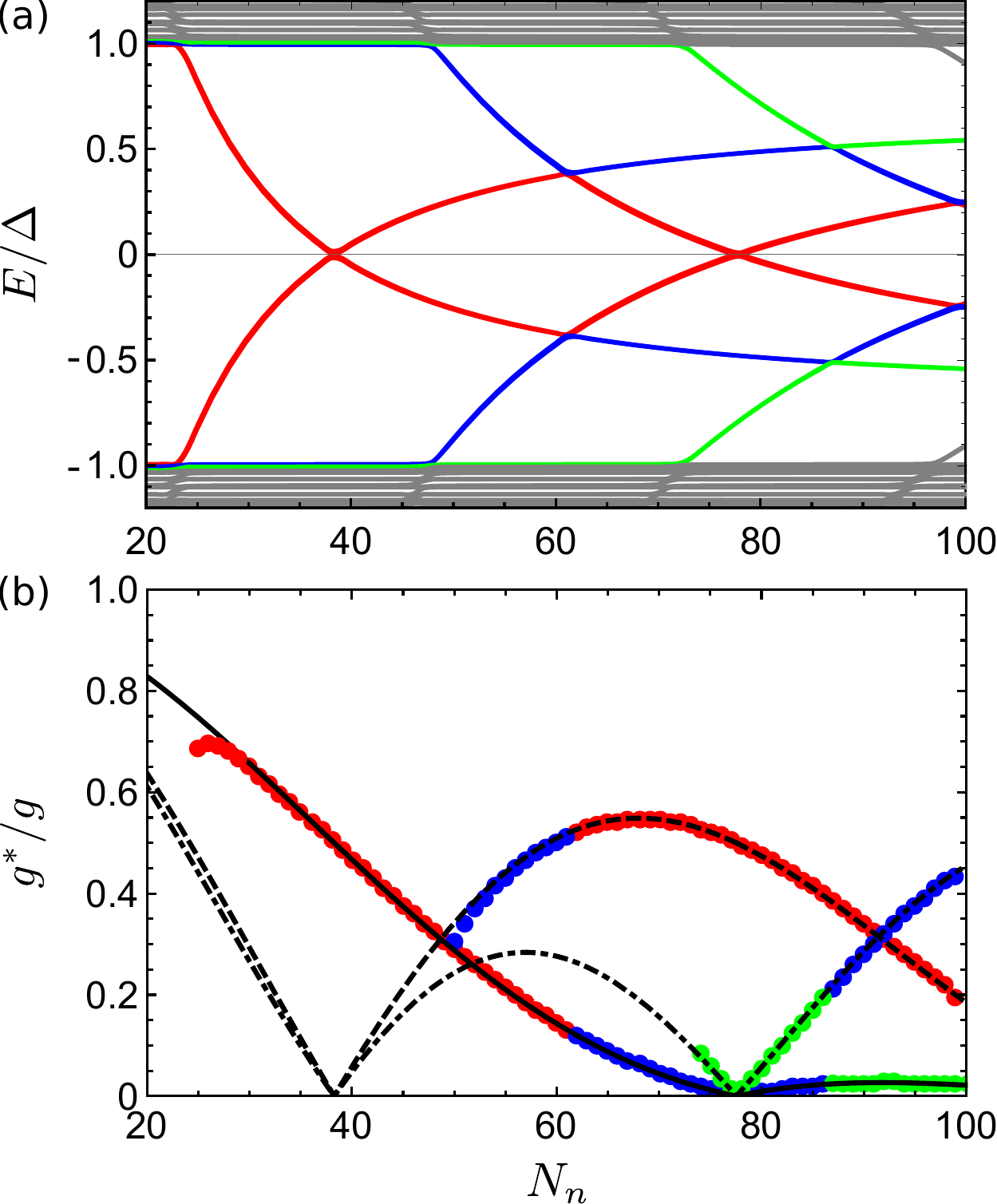}
\caption{(a) Energy spectrum, $E/\Delta$, and (b) effective $g$-factor, $g^*/g$, of the subgap states as a function of the dot size $N_n$ in the presence of a high ($\mu_b = -t$) and sharp ($W_b=1$)  barrier. In this configuration, we recover results obtained for an isolated QD with hard-wall confinement. The first ($m=1$, red), the second ($m=2$, blue), and  the third ($m=3$, green) dot level appear below the gap at different values of the dot size. Here, we label QD levels according to their energy distance to the chemical potential. (b) The effective $g$-factor, $(g^*/g)_m$, found numerically (colored dots) can be described well by analytical expressions obtained for hard-wall confinement [see Eq.~(\ref{eq:splittingn})], where we identify the three  lowest orbital levels of the QD with $n =1$ (solid black line), $n =2$ (dashed black line), and $n =3$ (dot-dashed black line). We note that numerically we can determine only the $g$-factors for the QD levels with energies below the  gap. 
 The parameters are chosen as  $\overline{\alpha} / t = 0.08$ ($k_{so} a = 0.08$), $\mu_s/t = \mu_n/t = -2$, $\Delta/t = 0.01$, and $V_Z/t=10^{-4}$.} 
\label{fig:Fig3}
\end{figure}

First, we check that in the case of a high barrier, we reproduce our previous results for the isolated QD with hard-wall confinement potential. Aiming at including later the superconducting section, the energy spectrum of the system is doubled due to the particle-hole symmetry. In addition, only QD levels lying inside the superconducting gap can be resolved. As a result, without fine-tuning,  there are no ABS levels in a small QD.
However, as the size of the QD is increased, more states enter in the superconducting gap as the level spacing scales  $\propto 1/L^2$ [see Fig.~\ref{fig:Fig3} (a)]. 
In the previous section, we have numbered the QD levels starting from the lowest orbital energy states at the bottom of the band. From now on,  we
will number the QD levels, which are two-fold degenerate at zero magnetic field, according to distance of their corresponding energies to the chemical potential which we fix to zero. The energy level closest to zero  is then labeled as  $m=1$ [depicted in red in Fig.~\ref{fig:Fig3} (a)], the second closest as $m=2$ [depicted in blue in Fig.~\ref{fig:Fig3} (a)], and the third closest as  $m=3$ [depicted in green in Fig.~\ref{fig:Fig3} (a)]. 
We note that if one follows the level labeled by $m=1$ continuously as the size of the QD increases, one is forced to jump from the $n=1$  to the $n=2$  level in the notation used in the previous section, where $n$ refers to the QD levels numbered from the bottom of the band. Working in this convention, we find that the effective $g$-factor in the case of high barrier is the same as the one found for the isolated normal QD, see Fig.~\ref{fig:Fig3} (b). Indeed, if the barrier is high,  QD levels are not subject to superconducting pairing, which is confirmed by the absence of anti-crossings between electron and hole levels.

\begin{figure}[t] 
\includegraphics[width=8cm]{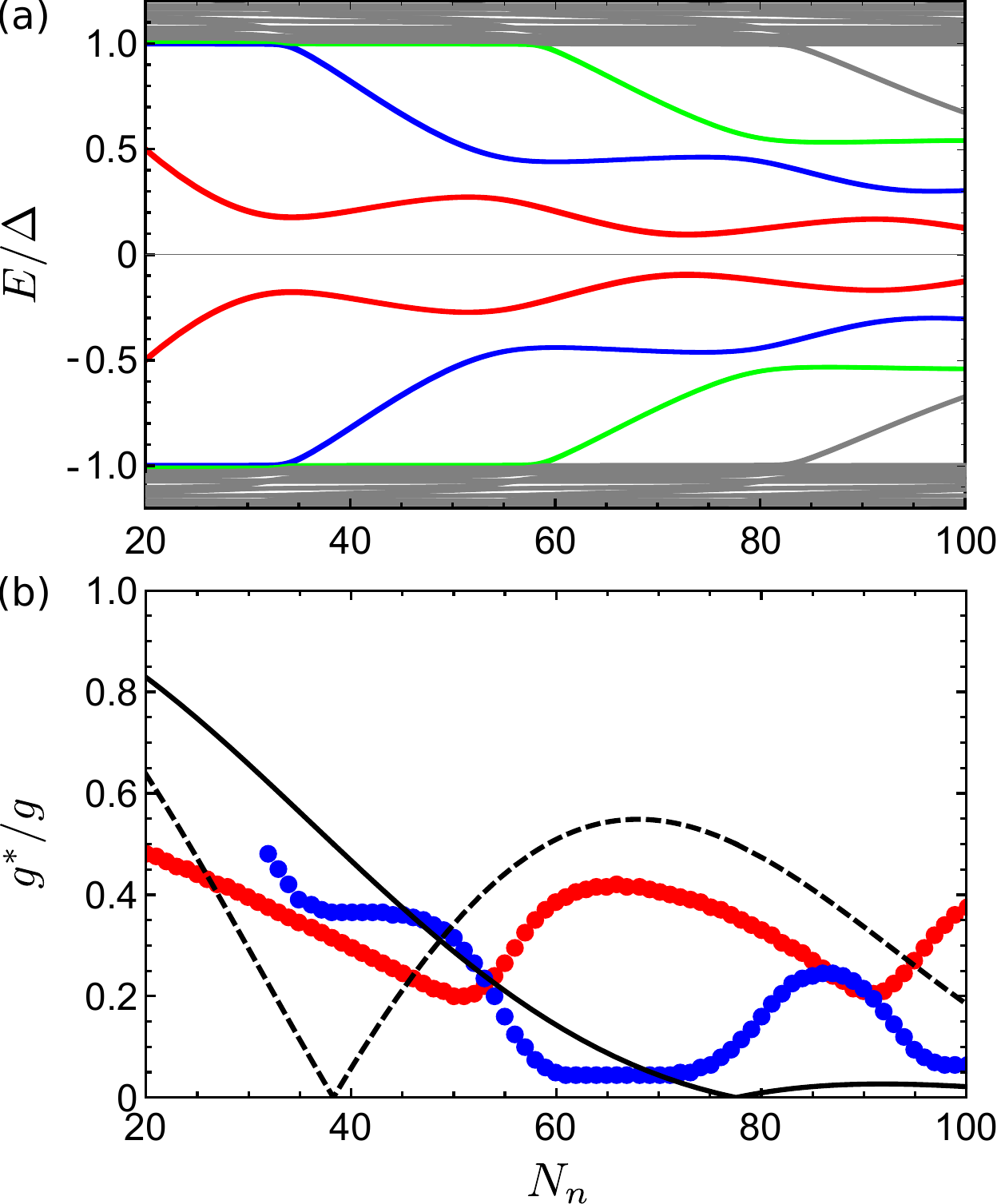}
\caption{The same as in Fig. \ref{fig:Fig3}, however, in the absence of a barrier, $\mu_b = 0$. (a) As the barrier is removed, the wavefunctions corresponding to the lowest QD levels become ABSs which extend also into the superconducting section. As a result, anti-crossings between electron and hole ABS levels are induced by the superconducting pairing. 
The renormalization of the effective $g$-factor is  still substantial, however, its functional form deviates from one obtained for an isolated normal QD. Generally, the $g$-factor suppression is less pronounced for large QDs.
} 
\label{fig:Fig4}
\end{figure}

\begin{figure}[t] 
\includegraphics[width=7.5cm]{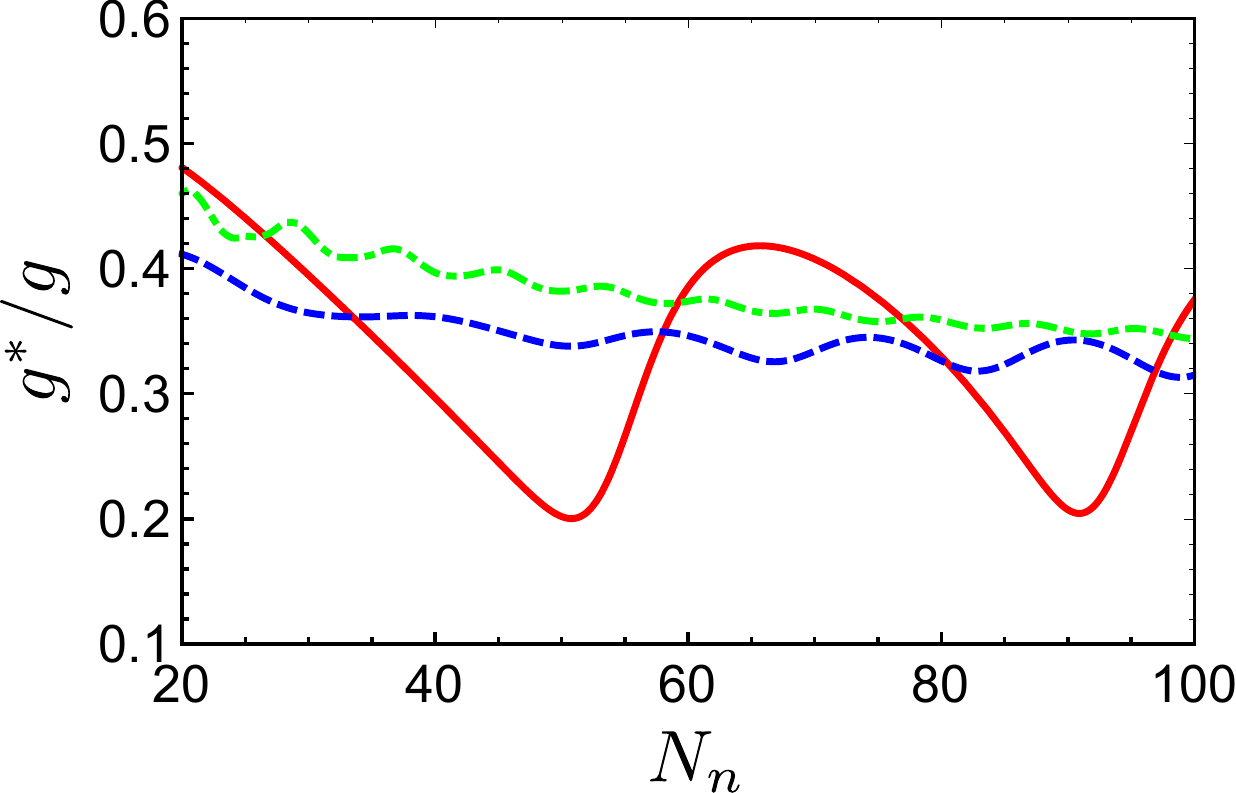}
\caption{Effective $g$-factor of the lowest ABS as a function of the dot size $N_n$  for different values of SOI: weak SOI regime with $\overline{\alpha}/t = 0.08$ (red solid line), moderate SOI regime with $\overline{\alpha}/t = 0.2$ (blue dashed line), and strong SOI regime with $\overline{\alpha}/t = 0.4$ (green dot-dashed line). The dependence on the SOI strength is less pronounced compared to the case of a normal isolated QD. The oscillations with the period set by $k_{so}$ are most pronounced in the weak SOI regime.
The parameters are chosen as  $\Delta/t = 0.01$, $\mu_n/t = \mu_s/t = -2$, and $V_Z/t = 10^{-4}$.
}
\label{fig:Figsoi}
\end{figure}

Next, we study how the presence of the superconducting section modifies the effective $g$-factor. In a first step, we completely remove the barrier at the NS interface, $\mu_b = 0$. As a result, the wavefunctions of the localized QD states penetrate into the superconducting section on the scale of the superconducting coherence length, which gives  rise to anticrossings between the particle and hole states [see Fig.~\ref{fig:Fig4}(a)]. The effective $g$-factor of such ABSs is again suppressed due to the presence of SOI, see Fig.~\ref{fig:Fig4}(b). However, compared to an isolated normal QD, the suppression is now stronger (weaker) for small (large) dots. Again, the $g$-factor suppression is governed by the strength of the SOI. In contrast to the normal QD, this dependence is much weaker, see Fig. \ref{fig:Figsoi}. Generally, the ratio $g^*/g$ is described by an oscillating function of $k_{so} L$. However, the oscillations are well-pronounced only in the regime of weak SOI.

\begin{figure}[b] 
\includegraphics[width=7.5cm]{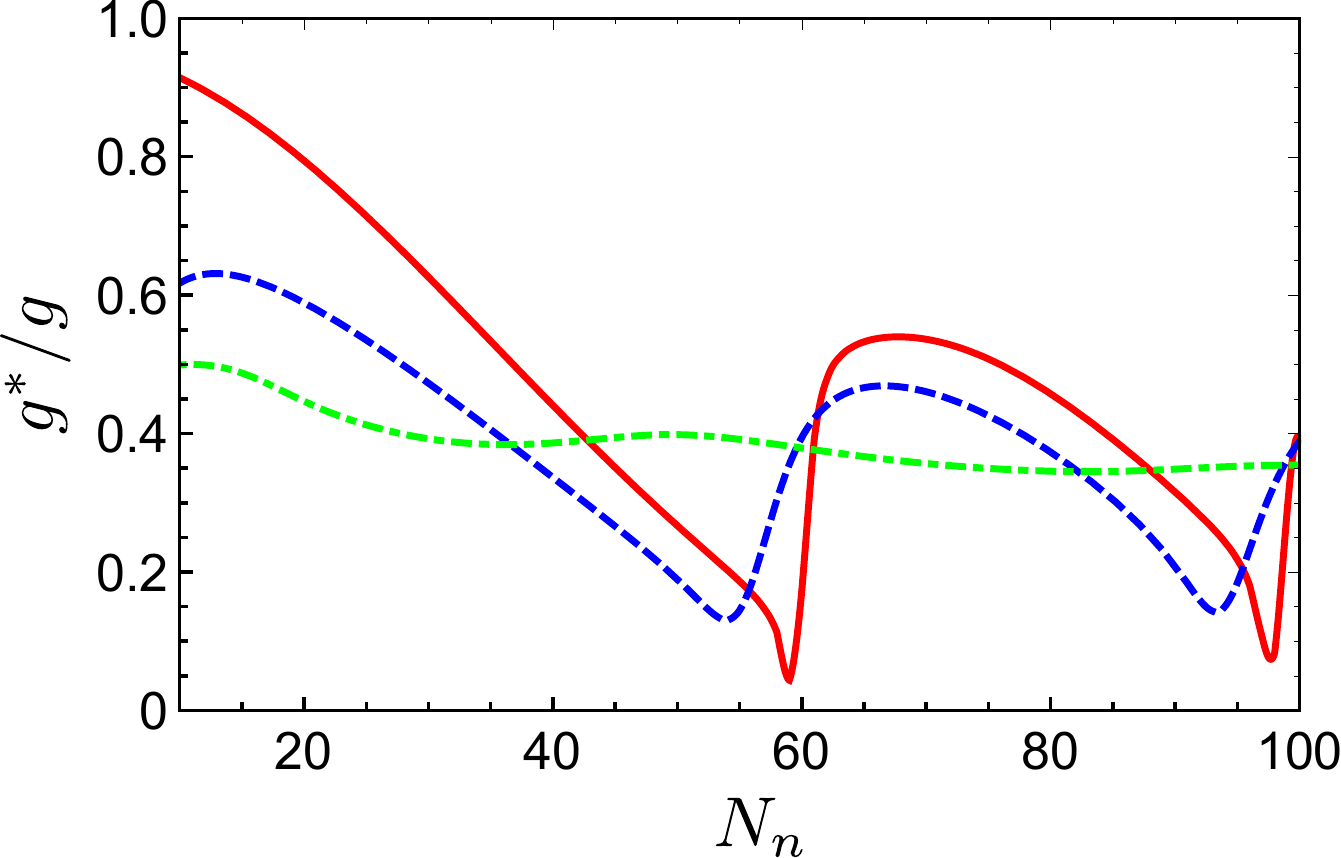}
\caption{Effective $g$-factor of the lowest ABS  as a function of dot size $N_n$ for different gap values:  $\Delta/t = 0.2$ (red solid line), $\Delta/t = 0.02$  (blue dashed line ), and  $\Delta/t = 0.002$ (green dot-dashed line). The oscillations of the effective $g$-factor almost disappear and $g^*/g$ becomes constant with decreasing $\Delta$. If the dot size  is much smaller than the SC localization length, the spatial support of the ABS is shifted to the superconducting section, and, thus, there is only a weak dependence on the parameters of the normal section.
The parameters are chosen as  $\overline{\alpha}/t = 0.08$, $\mu_n/t = \mu_s/t = -2$, and $V_Z/t = 10^{-4}$.
} 
\label{fig:Fig7}
\end{figure}

The $g$-factor suppression in the case of a normal isolated QD was solely described by the product of the SOI wavevector $k_{so}$ and the dot size $L$. In contrast to that, the $g$-factor suppression of the ABSs localized at the QD coupled to a superconducting section also depends on the size of the superconducting gap $\Delta$. In Fig. \ref{fig:Fig7}, $g^*/g$ is shown as a function of the QD size for different values of $\Delta$. If the gap is large, there are pronounced oscillations in the $g$-factor values. As $\Delta$ is decreased, the effective $g$-factor dependence on the dot size gets flattened and the oscillations disappear. This can be explained as follows. If the superconducting coherence length gets much larger than the size of the QD, the spatial support of the localized wavefunctions is shifted into the superconducting section, thus, the effective $g$-factor hardly depends on $L$.

\begin{figure}[b!] 
\includegraphics[width=7.5cm]{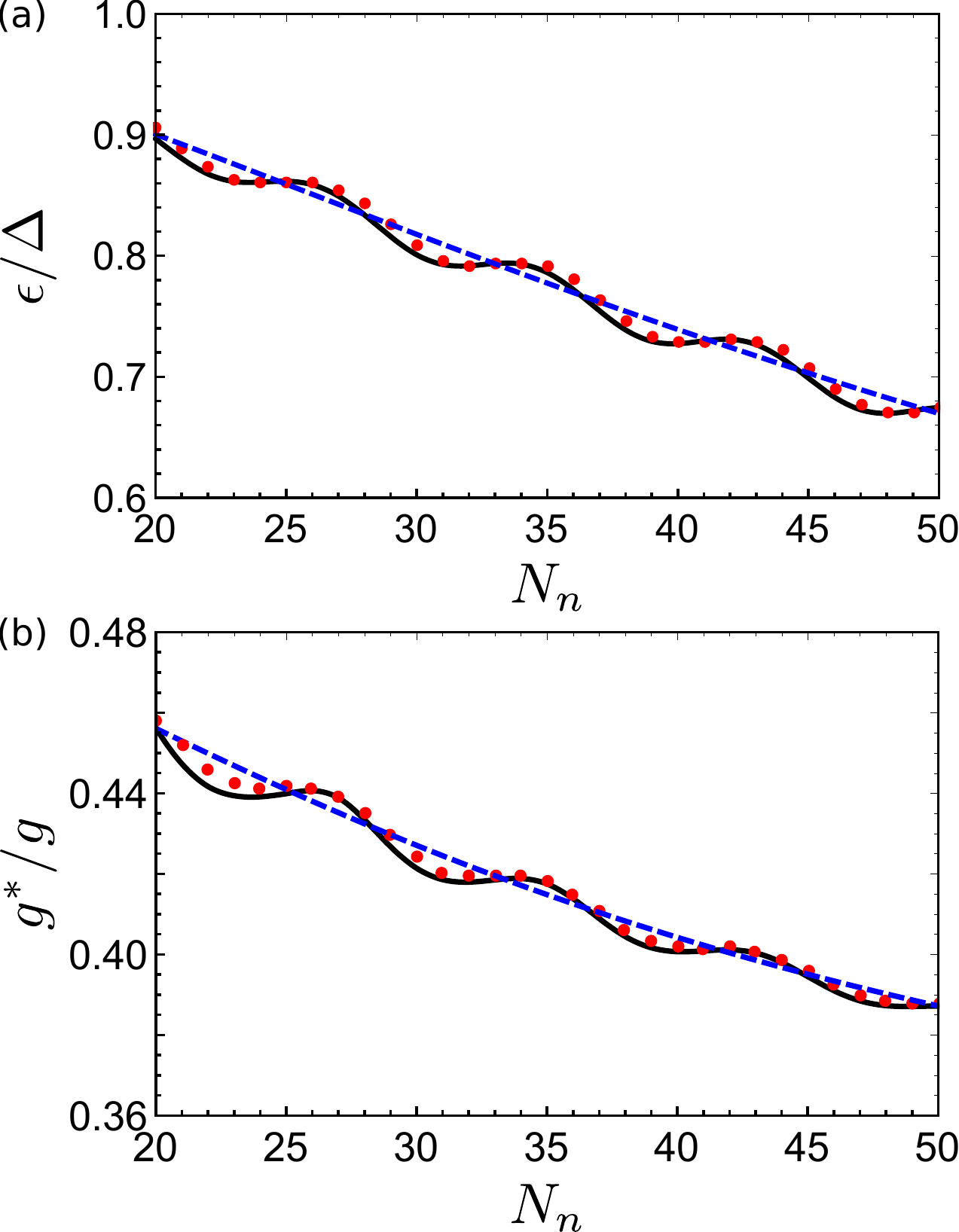}
\caption{(a) Energy  $\epsilon/\Delta$ of the lowest ABS  as  function of  dot size $N_n$ in the absence of  magnetic field, $V_Z = 0$. As  $N_n$ is increased, the bound state energy decreases and the ABS moves close to the chemical potential. The black line is obtained by numerical diagonalization of the tight-binding model [see  Eq. (\ref{eq:hamilbarr})], the red dotted line represents the numerical solution of the transcendental  equation for the energy spectrum obtained analytically [see  Eq. (\ref{eq:det1})], 
and the blue dashed line represents an approximate analytical solution [see Eq. (\ref{eq:epsilon})].  
(b) Effective $g$-factor of the lowest ABS  as a function of dot size $N_n$. The $g$-factor exhibits only weak dependence on $N_n$. For large QDs, it saturates to the length-independent value $g^*/g=1/\pi$.
The  black line  is obtained by numerical diagonalization of the tight-binding model [see  Eq. (\ref{eq:hamilbarr})], the red dotted line represents the numerical solution  obtained from the  transcendental  equation for the spectrum [see  Eq. (\ref{eq:det1}) and Eq. (\ref{eq:gfactorapp2})],
and the blue dashed line represents an approximate analytical solution [see Eq. (\ref{eq:epsilon}) and Eq. (\ref{eq:gfactormain})].
The parameters are chosen as $V_Z/t = 10^{-4}$, $\overline{\alpha}/t=0.4$, $\Delta/t=0.01$, $\mu_b = 0$, $\mu_s/t = \mu_n/t = -2$, and $N_s=500$.
} 
\label{fig:Fig9}
\end{figure}

The results presented above are obtained numerically. To get a better understanding of the observed behavior of the effective $g$-factor of ABSs localized at the transparent interface between normal and superconducting sections of the NW, we also study the setup analytically in the regime of strong SOI ($E_{so}\gg \Delta, V_Z$), in which the calculations can be performed by linearializing the spectrum around the two Fermi momenta,  see Ref. \onlinecite{dmytruk2018suppression} and Appendix \ref{sec:zeeman}. For the sake of simplicity, we assume that the chemical potential is uniform and is tuned to the SOI energy.

The spectrum of the ABSs $\epsilon$ is determined by solving the following transcendental equation,
\begin{align}
%\sin\left(2 \epsilon L / \alpha - \varphi\right) = 0,
2\epsilon L/\alpha -\arccos\left(\epsilon/\Delta\right) = \pi (p - 1).
\label{eq:epsilon}
\end{align}
In what follows, we focus only on states with positive energy $\epsilon > 0$. In this case, $p$ should be a positive integer.  First, by choosing $p=1$, we notice that Eq. (\ref{eq:epsilon}) has at least one solution, thus, there is always at least one ABS inside the superconducting gap. Generally, as the size of the QD, $L$, is increased, there are more and more ABSs. If $L_{p}<L<L_{p+1}$, there are $p$ ABSs, where we define $L_{p} = \alpha\left[\pi \left(p-1/2\right) - 1\right]/ 2\Delta$. The energy of these ABSs lying deeply inside the bulk gap can be approximated as 
\begin{align}
\epsilon_p \ \approx \pi \Delta\frac{p - 1/2}{1 + 2 L \Delta / \alpha}\, .
\label{eq:epsilonapprox}
\end{align} 
We note that our numerical results presented in Fig. \ref{fig:Fig9}a are in good agreement with our analytical formula for $\epsilon_1$ found from Eq. (\ref{eq:epsilon}). The level spacing $\delta$  is uniform and  given by $\delta/\Delta = \pi /\left(1 + 2 L \Delta / \alpha\right)$. As expected, it decreases as $1/L$ with increasing dot size.

Under the same approximations in the  strong SOI regime, the expression for the effective $g$-factor reads
\begin{align}
g^*/g \approx \dfrac{\alpha\Delta^2 \left|\cos \left(\frac{2 L \epsilon }{\alpha}\right)\right|}
{2 \epsilon^2\left(\alpha+2 L \sqrt{\Delta^2 -\epsilon^2}\right)},
\label{eq:gfactormain}
\end{align}
where $\epsilon$ is determined from Eq. (\ref{eq:epsilon}). The details of the derivation are given in Appendix~\ref{sec:zeeman}.  In Fig.~\ref{fig:Fig9}, we compare numerical (black line) and analytical (blue and red lines) results. There is a fairly good agreement between the two methods. The dependence of the $g$-factor on the QD size is weak and we do not observe the well-pronounced suppression that is typical for a normal QD, see in Sec.\ref{Parabolic/sharp}A.

In this work, we are predominantly interested in the effective $g$-factor of the lowest ABS ($p=1$) in the regime in which its energy $\epsilon_1$ is in the middle of the gap, $\epsilon_1/\Delta \ll 1$, which is the case for QDs of  size $L \gtrsim \alpha /2\Delta$. In this case, one can simplify Eq.~(\ref{eq:gfactormain}) as
\begin{align}
g^*/g = \frac{1}{\pi  \left[1-\pi ^2 \Delta  L/4\alpha(1 +2 \Delta  L/\alpha)^3\right]}.
\end{align}
We note that in the limit of large QDs, $L \gg \alpha /2\Delta$, the effective $g$-factor  saturates at a constant value, $g^*_1/g=1/\pi$, independent of the QD size. This is in strong contrast to the case of the QD confined by gates considered in Sec.~\ref{Parabolic/sharp}, where the $g$-factor was strongly suppressed and always went to zero as the size of the QD was increased, see Figs.~\ref{fig:Fig2b} and \ref{fig:Fig2a}. We note that a saturation to a length-independent  value of the $g$-factor $g_p^*$, $g^*_p/g \approx 1/[\pi\left(2 p -1\right)]$, is observed for all ABSs. The $g$-factor renormalization is stronger for high-energy levels, which is opposite to the results found for normal QDs.

\begin{figure}[ht] 
\includegraphics[width=7.5cm]{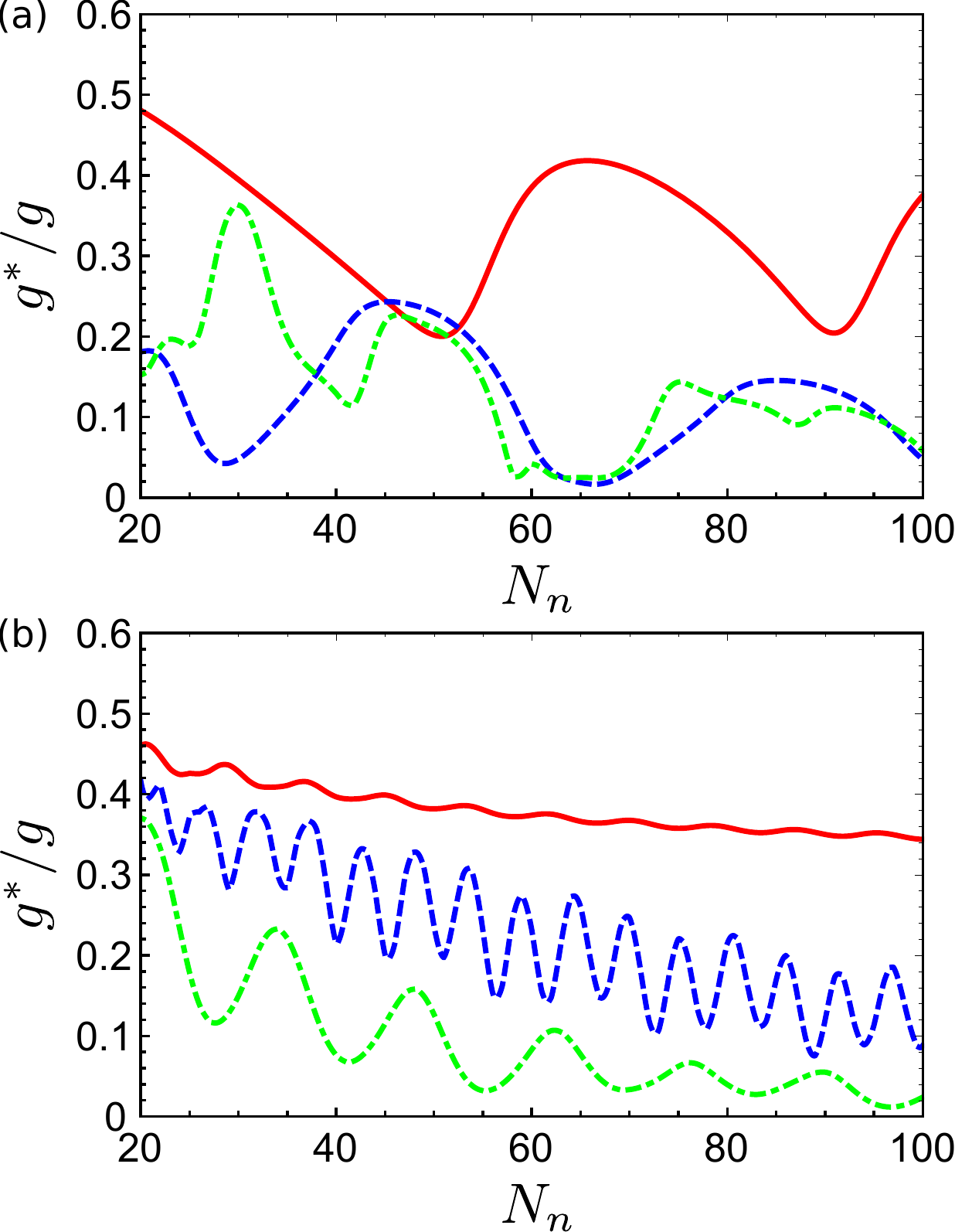}
\caption{Effective $g$-factor of the lowest ABS as  function of dot size $N_n$ for different values of $\mu_n$  (a) in weak SOI regime, $\overline{\alpha}/t = 0.08$, and (b)  in strong SOI regime, $\overline{\alpha}/t = 0.4$. In  panel (a) [(b)], red solid line corresponds to $\mu_n/t = -2$ ($\mu_n/t = -2$), blue dashed line  to $\mu_n/t = -1.98$ ($\mu_n/t = -1.8$), and green dot-dashed line  to $\mu_n/t = -1.96$ ($\mu_n/t = -2.1$). 
The other parameters are chosen as $\Delta/t = 0.01$, $\mu_s/t = -2$, and $V_Z/t = 10^{-4}$. In the strong SOI regime, the $g$-factor is maximal if the chemical potential is tuned to the SOI energy. Shifting away from this fine-tuned point results in strong suppression of $g^*$.}
%which can be detected in transport experiments.}
\label{fig:Figmu}
\end{figure}

To grasp the physics of the effects under consideration and to proceed with the analytical calculations of  the effective $g$-factor, we have assumed that the chemical potential is tuned to the SOI energy, corresponding to $\mu_n=\mu_s=-2t$. However, numerically, we can also study how $g^*$ is modified as we shift the chemical potential on the QD, see Fig.~\ref{fig:Figmu}. Generally, the $g$-factor is sensitive to the chemical potential position. Moreover, there is also a tendency for the $g$-factor to be at a maximum as the chemical potential is tuned to the SOI energy and to decrease as one tunes away from this fine-tuned value. In both weak and strong SOI regime, $g^*/g$ exhibits pronounced oscillations as a function of the QD size. The period of oscillations is set by the Fermi wavelength, such that the period is increasing (decreasing) as the density of electron in the NW decreases (increases). In addition, in the strong SOI regime, the amplitude of oscillation also grows as one detunes the chemical potential, however, it also decays as the size of the QD is increased. It is also important to note that, by changing the chemical potential on the QD, the $g$-factor can be tuned from some substantial finite value to almost zero, which opens up a way to control the Zeeman splitting of the ABSs  by changing solely the chemical potential while keeping the magnetic field fixed. This tunability is especially pronounced in the weak SOI regime. We note that all these features could be tested in transport experiments similar to the ones in Ref.~[\onlinecite{deng2016majorana}].

\begin{figure}[t] 
\includegraphics[width=7.5cm]{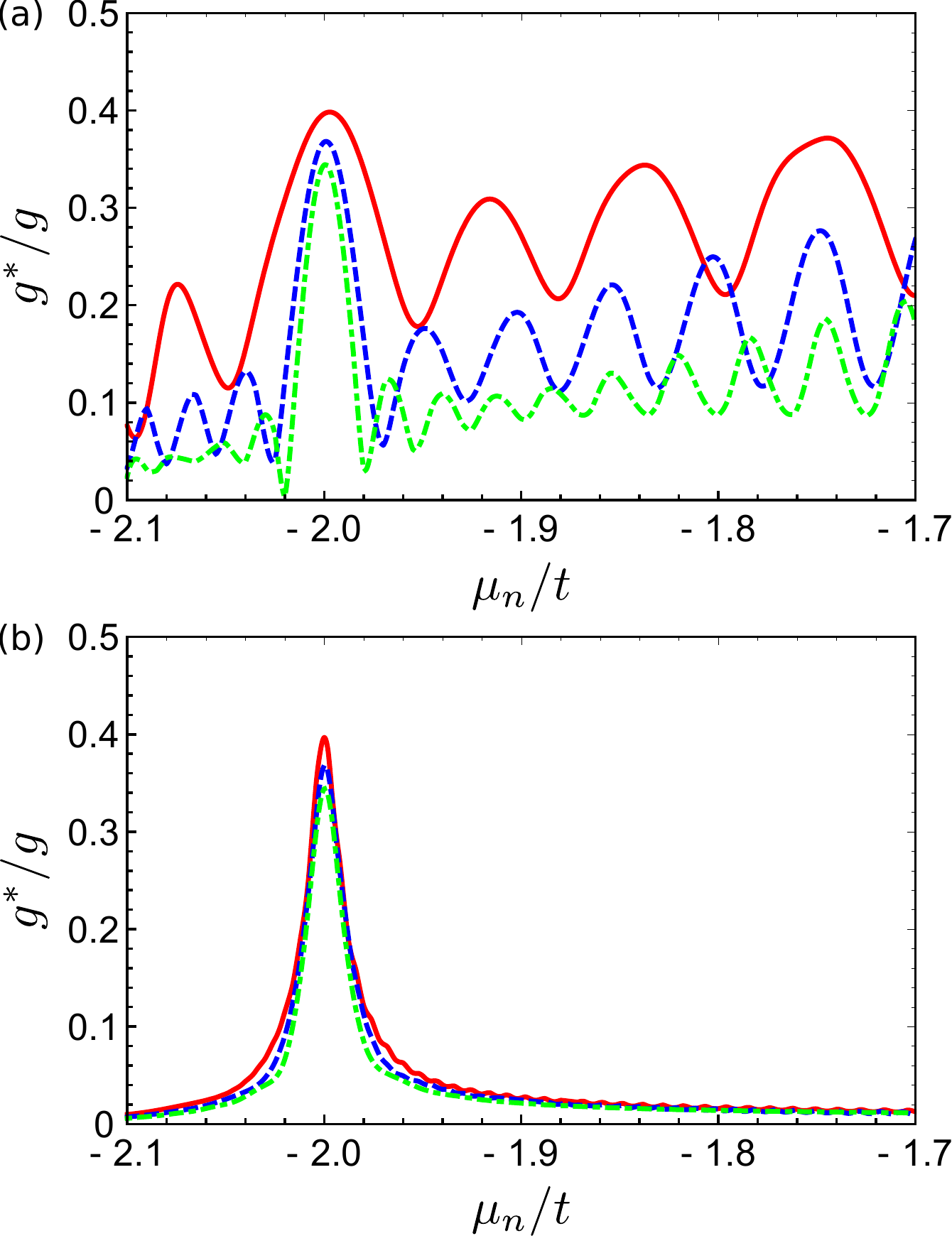}
\caption{Effective $g$-factor of the lowest ABS as  function of chemical potential $\mu_n/t$ in the normal section for different dot sizes:  $N_n=40$ (red solid line),  $N_n=70$ (blue dashed line), and  $N_n=100$ (green dot-dashed line). In  panel (a) the chemical potential in the SC section is fixed to the SOI energy $\mu_s/t = -2$, whereas in panel (b) we keep the chemical potential  uniform in the whole NW length, $\mu_n = \mu_s$. We note that in the latter case, the $g$-factor dependence on $\mu_n$ is described by the universal curve independent of the QD size.
In both panels, $g^*/g$ has a maximum at $\mu_n/t = -2$, corresponding to tuning the chemical potential to the SOI energy. 
The parameters are chosen as  $\overline{\alpha}/t=0.4$, $\Delta/t=0.01$, and $N_s=300$. }
\label{fig:FigN}
\end{figure}

Alternatively, if one detunes the position of the chemical potential in the superconducting section away from the SOI energy, $\mu_s= -2t$, however, again, one keeps the chemical potential in the normal section fixed to the SOI energy, $\mu_n = -2t$, we observe similar features, see Fig.\ref{fig:FigN}(a). The $g$-factor is maximal at $\mu_s= -2t$ and it gets smaller as one detunes the chemical potential. Again, there are oscillations set by the Fermi wavelength. However, in contrast to the previous case, in the strong SOI regime, the amplitude of oscillations grows with increasing electron density.

\begin{figure}[ht] 
\includegraphics[width=7.5cm]{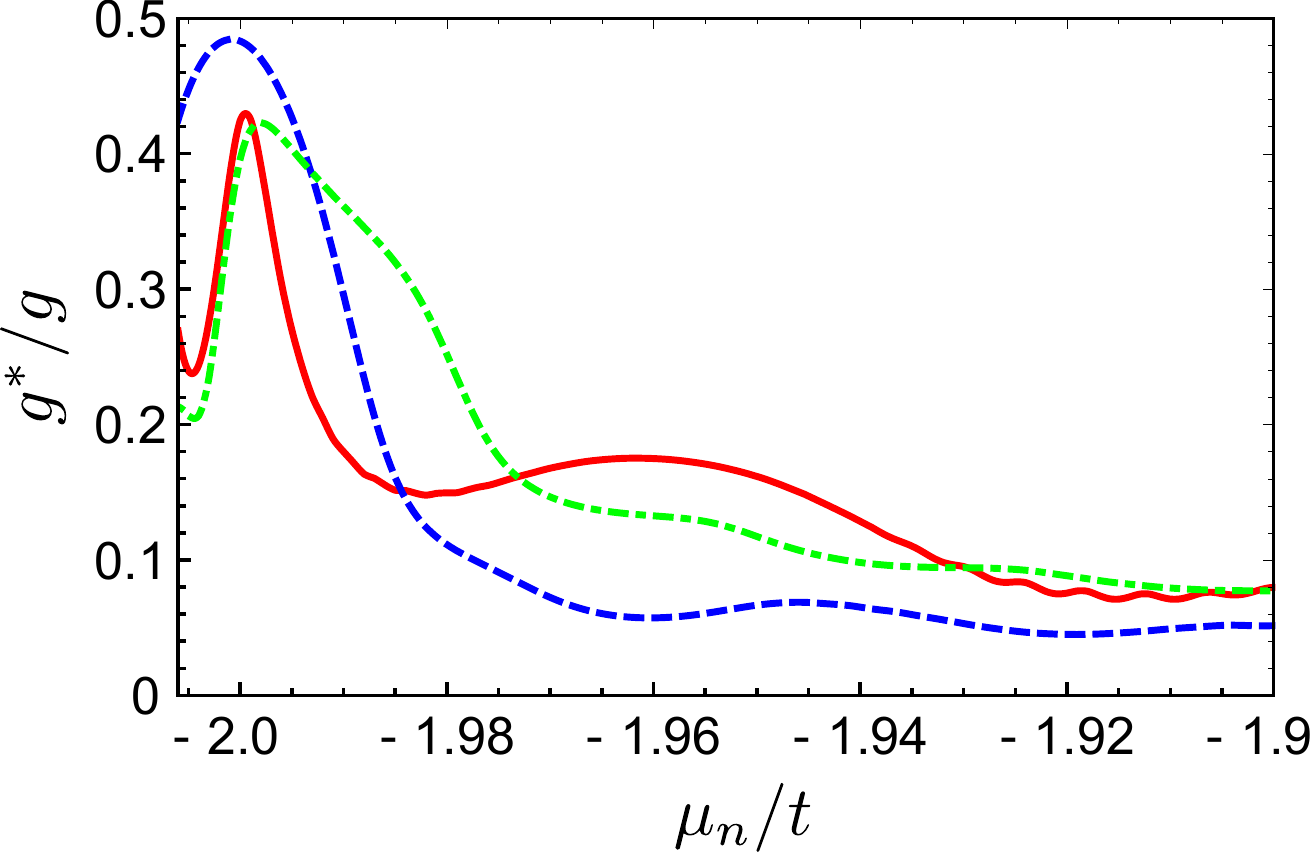}
\caption{Effective $g$-factor of the lowest ABS as  function  of  chemical potential $\mu_n = \mu_s$  for different dot sizes: $N_n=40$ (red solid line),  $N_n=70$ (blue dashed line), and  $N_n=100$ (green dot-dashed line).  The SOI is chosen to be stronger in the normal section, $\overline{\alpha}_n/t = 0.4$, than in the superconducting section, $\overline{\alpha}_s/t = 0.08$. Again, the $g$-factor reaches a maximum if both chemical potentials are tuned to their corresponding SOI energies.
The superconducting gap is fixed to  $\Delta/t = 0.01$.} 
\label{fig:Fig21}
\end{figure}

Next, we study the evolution of the effective $g$-factor of the lowest ABS located in the QD as a function of the chemical potentials in the normal and superconducting sections under the assumption that the chemical potential is shifted globally and both sections are kept at the same value, $\mu_n = \mu_s$. In this case, we see that $g^*/g$ has a well pronounced peak around $\mu_n =\mu_s= -2t$ and decays to zero otherwise [see Fig.\ref{fig:FigN} (b)]. Remarkably, the shape of the peak is  universal and does not depend on the length of the QD. In addition, we have checked that the same feature is observed even in the case when the SOI is non-uniform over the nanowire length, see Fig. \ref{fig:Fig21}. This is the case, for example, for Rashba SOI that can be strongly enhanced in the normal section by neighboring gates. Again, the renormalized $g$-factor is maximal as each of chemical potentials $\mu_s$ and $\mu_n$ is tuned to the SOI energy in the superconducting and normal sections, correspondingly. These findings provide a tool to access the SOI energy experimentally via measurements of the Zeeman splittings on the quantum dot. By tuning two gates, one acting in the normal section and the other in the superconducting section, one can determine the positions of the chemical potentials that correspond to the SOI energy. This can provide an important guidance for tuning such semiconducting-superconducting nanowires into the topological regime, where positioning of the chemical potential close to the SOI energy plays a crucial role.

\section{Conclusions}\label{conclusion} 
In this work, we have studied the renormalization of the $g$-factor of QD levels in the presence of Rashba SOI in a semiconducting-superconducting NW. For a QD  located in the normal section of the NW and separated from the superconducting section (with proximity-induced superconductivity) by high barrier, the effective $g$-factor depends only on the product of the SOI wavevector and the QD size, $k_{so} L$, and is strongly suppressed for the lowest orbital levels of the QD. If the QD confinement is harmonic, the $g$-factor is suppressed exponentially. In contrast to that, the hard-wall confinement results only in power-law suppression. In both setups, by changing the QD size, one can tune the effective $g$-factor to zero and, thus, one is able to suppress the Zeeman splitting completely. In addition, the $g$-factor as  function of $k_{so} L$ exhibits universal features that allow one to determine the quantum number of the dot level under  consideration.

As the barrier is lowered and the wavefunctions of the QD states leak into the superconducting section, the $g$-factor suppression is less pronounced. For example, in the regime of strong SOI, as the QD size is increased, the $g$-factor saturates to a constant value independent of $k_{so} L$ and of the superconducting gap $\Delta$. In small dots and in the weak SOI regime, the  $g$-factor exhibits strong oscillations as a function of the dot size. In the strong SOI regime,  such oscillations are suppressed. We predict that the $g$-factor is maximal if the chemical potential is tuned to the SOI energy in both sections of the NW. This remarkable feature can be used to tune the position of the chemical potential to the SOI energy in experiments.

\emph{Acknowledgments.} We acknowledge helpful discussion with Pavel Aseev. This work was supported by the Swiss National Science Foundation and the NCCR QSIT. This project received funding from the European Union’s Horizon 2020 research and innovation program (ERC Starting Grant, grant agreement  No 757725). 

\appendix

\section{Renormalization of the $g$-factor in strong SOI regime} 
\label{sec:zeeman}

In this section, for simplicity, to proceed with analytical calculations, we assume that the SOI is uniform in the entire nanowire and the chemical potential is tuned to the SOI energy. We again apply perturbation theory  and first find the energy spectrum of the ABSs as well as the corresponding wavefunctions in the absence of magnetic fields.
 To simplify our calculations, we can gauge away the SOI term by the transformation defined in Eq. (\ref{eq:spingaugetransform}).
In this case, we can linearilize the spectrum around two Fermi points $\pm k_{so}$. In addition, we express the electron operators $\tilde{\psi}_\sigma$ in terms of slowly varying left ($\tilde{L}_\sigma$) and right ($\tilde{R}_\sigma$) movers, $\tilde{\psi}_\sigma (x) = \tilde{R}_\sigma(x) e^{i k_{so}x} + \tilde{L}_\sigma(x) e^{-i k_{so}x}$.

As a result, the kinetic term in the Hamiltonian is given by
\begin{align}
&\tilde{H}_{kin}= -i \alpha \sum_{\sigma}\int_{0}^\infty dx  \left[
\tilde{R}^\dag_{\sigma}(x) \partial_x \tilde{R}_{\sigma}(x) - \tilde{L}^\dag_{\sigma}(x) \partial_x \tilde{L}_{\sigma}(x)\right],
\end{align}
where the SOI parameter $\alpha \geq 0  $ determines the Fermi velocity $\upsilon_F = \alpha/\hbar$.
The superconducting term in the Hamiltonian is given by
\begin{align}
\tilde{H}_{sc} = \Delta\int_L^\infty dx \ \left[\tilde{R}_{\uparrow}(x)\tilde{L}_{\downarrow}(x) + \tilde{L}_{\uparrow}(x)\tilde{R}_{\downarrow}(x)+\text{H.c.}\right].
\end{align}

Here, we focus exclusively on the ABSs below the SC gap and assume that the nanowire is long enough to avoid any boundary effects on the right end of the nanowire. This allows us to work in the limit of semi-infinite nanowires and impose the boundary conditions only at $x=0$ and $x=L$. To simplify our calculations further, we also assume that the barrier is absent, which corresponds to a transparent interface between normal and superconducting sections. In addition, we note that our system is time-reversal invariant and spin is a good quantum number. As a result, our Hamiltonian is block-diagonal in the spin degrees of freedom and all ABSs are twofold degenerate. This fact allows us, first, to focus on one of two ABSs and find  its wavefunction $\tilde{\Phi}_{1}$. The wavefunction $\tilde{\Phi}_{\bar1}$ of its Kramers partner is then easily found with the help of the time-reversal operator, $T= -i \sigma_y K$, $\tilde{\Phi}_{\bar1} = T \tilde{\Phi}_{1}$. Here, $K$ denotes the complex conjugation operator.

First, we find the two-component wavefunction $\phi_{1}$ of the Hamiltonian written in the basis  $\left(\tilde{\psi}_\uparrow, \tilde{\psi}^\dag_\downarrow\right)$.
The wavefunction of an ABS at a given energy $\epsilon$ below the superconducting gap, $|\epsilon|<\Delta$, is defined piecewise
\begin{equation}
\phi_{1} (x) = \begin{pmatrix}
f(x)\\
g(x)
\end{pmatrix}
=
\begin{cases}
a_1 \phi_n^{(1)}(x) +a_2 \phi^{(2)}_n(x), & 0\leq x \leq L,\\
b_1  \phi_s^{(1)}(x) + b_2 \phi_s^{(2)}(x), & x>L,
\end{cases},
\end{equation}
where  $\phi_n^{(1,2)}(x)$ are chosen such that they satisfy  vanishing boundary conditions at $x=0$,
\begin{align}
&\phi_n^{(1)}(x) = 
\begin{pmatrix}
\sin\left[\left(k_{so}-\epsilon/\alpha\right)x\right]\\
0
\end{pmatrix},\\
&\phi_n^{(2)}(x) = 
\begin{pmatrix}
0\\
\sin\left[\left(k_{so}+\epsilon/\alpha\right)x\right]
\end{pmatrix}.
\end{align}
In the superconducting section, we take into account only decaying eigenfunctions of the Hamiltonian
\begin{align}
&\phi_s^{(1)}(x) = e^{-\sqrt{\Delta^2-\epsilon^2}x / \alpha}
\begin{pmatrix}
\cos(k_{so} x - \varphi) \\
-\cos(k_{so} x)
\end{pmatrix},\\
&\phi_s^{(2)}(x) = e^{-\sqrt{\Delta^2-\epsilon^2}x / \alpha}
\begin{pmatrix}
\sin(k_{so} x - \varphi)\\
-\sin(k_{so} x)
\end{pmatrix},
\end{align}
where $\cos\varphi = \epsilon/\Delta$.
The set of  coefficients $\{a_1,a_2,b_1,b_2\}$ is determined by the boundary conditions at $x=L$,
 \begin{align}
&\phi_{1}(x = L^-) = \phi_{1}(x = L^+),\label{eq:boundarycondition1}\\
&\partial_x \phi_{1}(x = L^-) = \partial_x \phi_{1}(x = L^+).
\label{eq:boundarycondition2}
\end{align}
In addition, the wavefunction should be normalized, $\int_{0}^{+\infty}dx \left|\phi_1(x)\right|^2=1$.  All these conditions together allow us to determine the energy of the ABSs and its wavefunction.
The energy spectrum  $\epsilon$ is determined by solving the following equation: 
\begin{align}
&2 \alpha  k_{so} \cos \varphi  \Big[\alpha  k_{so} \sin \left(2 L \epsilon /\alpha \right)-\epsilon  \sin (2 k_{so} L)\Big]\nonumber\\
&+\sin \varphi  \Big\{-\left(\Delta ^2+2 \alpha ^2 k_{so}^2-2 \epsilon ^2\right) \cos \left(2 L \epsilon /\alpha \right)\nonumber\\
&+2 \sqrt{\Delta ^2-\epsilon ^2} \left[\epsilon  \sin \left(2 L \epsilon /\alpha \right)-\alpha  k_{so} \sin (2 k_{so} L)\right]\nonumber\\
&+\Delta ^2 \cos (2 k_{so} L)\Big\} = 0.
\label{eq:det1}
\end{align}
In the limit of strong SOI, $E_{so} \gg  \Delta> |\epsilon|$, this complicated equation in $\epsilon$ can be substantially simplified, resulting in $\sin\left(2 \epsilon L / \alpha - \varphi\right) = 0$.
Using the boundary conditions given by Eqs.(\ref{eq:boundarycondition1})and (\ref{eq:boundarycondition2}), we also find the set of coefficients, 
\begin{align}
&a_{1}/a_2  \approx -1, \\
&b_{1}/a_2 \approx  -  e^{\frac{L \sqrt{\Delta ^2-\epsilon ^2}}{\alpha}} \sin \left(L \epsilon /\alpha\right),\\
&b_{2}/a_2 
\approx -e^{\frac{L \sqrt{\Delta ^2-\epsilon ^2}}{\alpha}} \cos \left(L \epsilon /\alpha\right).
\end{align}
Using the normalization condition imposed on the wavefunction, we determine $a_2$ as $a_2=1/\sqrt{\mathcal{N}}$ with
\begin{align}
\mathcal{N}\approx L + \frac{\alpha}{2\sqrt{\Delta ^2-\epsilon ^2}}.
\end{align}

In the original basis,
the wavefunctions $\Phi_{1,\bar 1}$ that correspond to two degenerate ABSs read as
\begin{align}
&\Phi_1(x) =
\begin{pmatrix}
f(x) e^{-i k_{so}x}\\
0\\
0\\
g(x) e^{-i k_{so}x}
\end{pmatrix},\\
&\Phi_{\bar 1}(x) =
\begin{pmatrix}
0\\
f^*(x) e^{i k_{so} x}\\
-g^*(x) e^{i k_{so} x}\\
0
\end{pmatrix}.
\end{align}
Treating the Zeeman term $\mathcal{H}_Z = V_Z \sigma_x \tau_z$ as a perturbation, we calculate the effective $g$-factor using the same procedure as in Sec.~\ref{Parabolic/sharp}. 
Here, $\tau_z$ is the Pauli matrix acting in particle-hole space. 
 Using  degenerate perturbation theory, we find that the effective $g$-factor is given by
\begin{align}
g^*/g = \left|\int_{0}^{+\infty} dx\ e^{2 i k_{so} x} \left[f^2(x) + g^2(x)\right]\right|.
\end{align}
In the strong SOI regime, we find that the effective $g$-factor reads
\begin{align}
g^*/g \approx  \dfrac{\alpha}{4\mathcal{N}}\left|\frac{\sin \left(\frac{2 L \epsilon }{\alpha }\right)}{\epsilon } +\left(1 + e^{2 i \varphi}\right) \frac{e^{-2 i \epsilon L/\alpha}}{2 \sqrt{\Delta ^2-\epsilon ^2}} \right|.
\label{eq:gfactorapp}
\end{align}
Using the approximate equation for $\epsilon$, $\sin\left(2 \epsilon L / \alpha - \varphi\right) = 0$, we can see that the imaginary part within the brackets vanishes (it is small for the exact value of $\epsilon$ found from Eq.(\ref{eq:det1})), and $g^*/g$ is given by 
\begin{align}
&g^*/g \approx \nonumber\\
&\dfrac{\alpha\left|\left(\Delta ^2+\epsilon ^2\right) \sqrt{\Delta^2 -\epsilon^2 } \sin \left(\frac{2 L \epsilon }{\alpha}\right)+\epsilon ^3 \cos \left(\frac{2 L \epsilon }{\alpha}\right)\right|}
{2 \Delta ^2 \epsilon\left(\alpha+2 L \sqrt{\Delta^2 -\epsilon^2}\right)}
.
\label{eq:gfactorapp2}
\end{align}
To further simplify the expression for the effective $g$-factor, we use the approximate equation for $\epsilon$ arriving at 
\begin{align}
g^*/g \approx \dfrac{\alpha\Delta^2 \left|\cos \left(\frac{2 L \epsilon }{\alpha}\right)\right|}
{2 \epsilon^2\left(\alpha+2 L \sqrt{\Delta^2 -\epsilon^2}\right)}
.
\label{eq:gfactorapp3}
\end{align}
This is the approximate expression for $g^*/g$ displayed in the main text. It captures the main features of the Zeeman splitting between the two lowest energy states.

\end{document}